\documentclass[twocolumn,amsmath,amssymb,superscriptaddress]{revtex4}
\usepackage[bookmarks,linktocpage, colorlinks=true, plainpages = false, citecolor = blue,  linkcolor=blue, urlcolor = blue,
filecolor = blue]{hyperref}

\usepackage{graphicx}
\usepackage{dcolumn}
\usepackage{bm}
\usepackage{braket}

\input epsf
\def\gsim{ \lower .75ex \hbox{$\sim$} \llap{\raise .27ex \hbox{$>$}} }
\def\lsim{ \lower .75ex \hbox{$\sim$} \llap{\raise .27ex \hbox{$<$}} }

\newcommand{\be}{\begin{equation}}
\newcommand{\ee}{\end{equation}}
\newcommand{\bea}{\begin{eqnarray}}
\newcommand{\eea}{\end{eqnarray}}

\def\de{\delta}

\def\ep{\epsilon}

\def\m{\mu}
\def\n{\nu}

\def\t{\tau}

\def\z{\zeta}

\def\p{\partial}

\newcommand{\ba}{\begin{aligned}}
\newcommand{\ea}{\end{aligned}}

\newcommand{\cL}{\mathcal{L}}

\newcommand{\fd}{\mathrm{d}}
\newcommand{\cH}{\mathcal{H}}
\newcommand{\h}{\frac{1}{2}}

\renewcommand{\t}{\tau}

\begin{document}
\begin{titlepage}

\title{Ekpyrotic Perturbations With Small Non-Gaussian Corrections}

\author{Angelika Fertig}
\author{Jean-Luc Lehners}
\author{Enno Mallwitz}

\affiliation{Max--Planck--Institute for Gravitational Physics (Albert--Einstein--Institute), 14476 Potsdam, Germany}

\begin{abstract}

\noindent The entropic mechanism for producing nearly scale-invariant density perturbations in a contracting ekpyrotic universe relies on having an unstable scalar potential. Here we develop a variant of this mechanism (recently proposed by Qiu, Gao and Saridakis, and by Li), in which there exists a non-trivial coupling between adiabatic and entropic fields, and where an unstable potential is not required. In the model nearly scale-invariant entropy perturbations are generated first. Remarkably, we find that the bispectrum of these perturbations vanishes, with the values of the non-Gaussianity parameters of local, equilateral and orthogonal type all exactly zero. Subsequently, the entropy perturbations can be converted into curvature perturbations by a variety of mechanisms. The bispectrum of the curvature perturbations depends on the non-linearity of the conversion process and is thus more model-dependent -- however, for an efficient conversion process the final bispectrum remains small. The only distinguishing feature compared to single-field slow-roll inflationary models is an absence of primordial gravitational waves. Thus the present model provides a perfect match to current data from the PLANCK satellite.

\end{abstract}
\maketitle

\end{titlepage}

Recent measurements by the PLANCK satellite show the early universe to have been extraordinarily simple \cite{Ade:2013zuv,Ade:2013uln,Ade:2013ydc}: not only approximately flat, homogeneous and isotropic, but also containing nearly scale-invariant and Gaussian density fluctuations. A major goal of cosmology is to find a convincing explanation for this initial state. The most popular current model is that of single-scalar-field inflation with a flat plateau-like potential, and the predictions of such a model are indeed in good agreement with observations. However, from a theoretical point of view, plateau models of inflation present significant challenges (see e.g. \cite{Ijjas:2013vea}), the most important one being perhaps that they typically lead to the runaway behavior of eternal inflation \cite{Guth:2007ng}: the implied infinity of disconnected and physically different universes causes the theory to lose all its predictive power and renders its naive predictions questionable. 

This situation suggests two complementary approaches: the first is to try to resolve the inflationary challenges. The second is to look for alternative theories which might be able to explain the same cosmological data without however presenting us with conceptual conundrums. In the present paper we will be concerned with the second approach.

An attractive alternative to inflation are ekpyrotic/cyclic models of the universe \cite{Khoury:2001wf,Steinhardt:2001st,Lehners:2008vx}. In these models the universe alternates between contracting and expanding phases, with the Big Bang corresponding to the reversal from contraction to expansion. The cosmological flatness problem can be solved during the contracting phase so long as the equation of state of the universe, $w=p/\rho$ (with $p$ being the pressure and $\rho$ the energy density), is larger than $1$. This can be achieved by having a scalar field $\phi$ with a steep and negative potential. Here we will consider potentials that are of exponential form,
\be \label{eq:ekpot}
V(\phi) = - V_0 e^{-c \phi},
\ee
and the requirement $w>1$ is then equivalent to $c>\sqrt{6}.$ In the presence of a second scalar field $\chi$ with a canonical kinetic term and an {\it unstable} direction in the potential $V_{,\chi\chi}<0,$ nearly scale-invariant (quantum) entropy fluctuations get amplified to become classical perturbations, which can then be converted into curvature perturbations before the Big Bang \cite{Notari:2002yc,Finelli:2002we,Lehners:2007ac,Battarra:2013cha}. In this ``entropic mechanism'' the spectrum is predicted to be close to scale-invariant, while the local bispectrum can take on a range of values \cite{Lehners:2007wc,Lehners:2009ja,Lehners:2010fy}, including the values favored by PLANCK \cite{Ade:2013ydc,Lehners:2013cka}. The fact that an unstable potential is required (the consequences of which are discussed in \cite{Lehners:2009eg}) is perceived by many as the weakest point of the model \footnote{In order to obtain a complete history from the ekpyrotic phase until the present one must supplement these models with a description of the bounce. There exist two broad classes of bounce models: in the first the scale factor reaches zero but quantum and/or higher-dimensional effects are conjectured to resolve this singularity (see e.g. \cite{Turok:2004gb,Bars:2011aa}), and in the second the scale factor reaches a minimum at a non-zero value so that the evolution can be described classically throughout (see e.g. \cite{Buchbinder:2007ad,Creminelli:2007aq,Lehners:2011kr,Easson:2011zy,Cai:2012va,Qiu:2013eoa}). It has recently been demonstrated that such non-singular bounces are also viable in supergravity \cite{Koehn:2013}.}.

\indent {\it The Model.} In the present paper we are interested in a new mechanism for generating ekpyrotic density perturbations. This mechanism was proposed by Qiu, Gao and Saridakis \cite{Qiu:2013eoa} as well as by Li \cite{Li:2013hga}, and it has the advantage that it does not require an unstable potential. In fact, in this new model, which we will refer to as the ``non-minimal entropic mechanism'', the potential does not depend on the second scalar $\chi$ at all, but one introduces a non-minimal coupling between the scalars $\phi$ and $\chi$ in the kinetic term for $\chi,$ by considering the Lagrangian
(in natural units $8 \pi G=M_{Pl}^{-2}=1$)
\be \label{eq:Lagrangian}
\cL =\sqrt{-g}\left[ \frac{R}{2} -  \h \p_\m \phi \p^\m \phi - \h e^{- b \phi} \p_\m \chi \p^\m \chi + V_0 e^{-c \phi}  \right],
\ee
where for now we assume $b,c$ to be constants. In a flat Friedmann--Lema$\hat{\imath}$tre-Robertson-Walker (FLRW) universe, with metric $\fd s^2 = - \fd t^2 + a(t)^2 \delta_{ij} \fd x^i \fd x^j,$ where $a(t)$ is the scale factor and with $\dot{}\equiv \frac{\fd}{\fd t},$ the equations of motion are given by
\bea \label{eq:phieom}
&& \ddot{\phi} + 3 H \dot{\phi} + cV_0 e^{-c \phi} = - \h b e^{- b \phi} \dot{\chi}^2, \\
&& \ddot{\chi} + \left( 3 H -b \dot{\phi} \right) \dot{\chi} + e^{b \phi} V_{, \chi} = 0,  \label{eq:chieom} \\
&& H^2 = \frac{1}{6} \left(\dot{\phi}^2 + e^{- b \phi} \dot{\chi}^2 - 2 V_0 e^{-c\phi} \right).  \label{eq:fried1}
\eea
Since the potential $V(\phi)$ does not depend on $\chi,$ it is clear that $\chi = constant$ is a solution. The remaining equations then reduce to those for a single scalar in an ekpyrotic potential, and they admit the scaling solution \cite{Khoury:2001wf}
\be
a \propto (-t)^{1/\ep}, \, \phi = 
\sqrt{\frac{2}{\ep}} \ln{\left[ - \left( \frac{V_0 \ep^2}{ (\ep -3)} \right)^{\h}  t\right]}, \, \ep=\frac{c^2}{2},
\ee
where $t$ is negative and runs from large negative towards small negative values. Here $\ep \equiv  \dot\phi^2/(2H^2)$ is directly related to the equation of state $w=2\ep/3-1$ and hence $\ep>3$. A standard analysis shows that in order to solve the flatness problem $|a H|$ has to grow by at least $60$ e-folds over the course of the ekpyrotic phase. Using Eq. (\ref{eq:phieom}), this implies a minimum field range $|\Delta \phi| > 60\sqrt{2\ep}/(\ep-1).$ It is over this field range that the potential must take the form expressed in Eq. \eqref{eq:ekpot}. As is intuitively clear, the steeper the potential, the shorter the required field range.

\indent {\it Entropy Perturbations.} Having discussed the background, we now turn our attention to the fluctuations. The main source of perturbations in the non-minimal entropic model are the entropy fluctuations. These correspond to fluctuations that are transverse (in scalar field space) to the background trajectory. A useful definition for the entropy perturbation is given by the gauge-invariant quantity $\de s = e^{- \frac{b}{2} \phi}  \de \chi,$ whose linearized equation of motion in Fourier space is given by \cite{DiMarco:2002eb}
\be \label{eq:deltaseom_2}
\ddot{\de s} + 3 H \dot{\de s} + \left[ \frac{k^2}{a^2}  - \frac{b^2}{4} \dot{\phi}^2   - \frac{b}{2}  V_{,\phi}  \right]  \de s =0,
\ee
where $k$ denotes the wavenumber of the fluctuation mode. Switching to conformal time, defined via $\fd t = a \fd \tau,$ with $'\equiv \frac{\fd}{\fd \t},$ and defining the canonically normalized entropy perturbation $v_s \equiv a \de s,$ we obtain
\be \label{eq:deltaseomtau_2}
v_s''  + \left[ k^2 - \frac{a''}{a}  - \frac{b^2}{4}\phi'^2  - \frac{b}{2} a^2  V_{,\phi}  \right] v_s =0.
\ee
Imposing the usual boundary condition that in the far past/on small scales the mode function is that of a fluctuation in Minkowski space, $\lim\limits_{k\tau \rightarrow -\infty}{v_s} = \frac{1}{\sqrt{2 k}} e^{- i k \tau},$ up to an irrelevant phase the solution is
\be
v_s = \sqrt{\frac{\pi}{4}} \sqrt{-\t} H_\nu^{(1)}(-k\t),
\ee
where $H_\n^{(1)}$ denotes a Hankel function of the first kind. At late times/on large scales, the entropy perturbations then scale as
\be \label{eq:modefunctionlate}
v_s \propto k^{-\nu} (-\t)^{1/2-\nu}  \qquad (|k\t| \ll 1).
\ee
Defining a parameter $\Delta \equiv \frac{b}{c}-1,$ 
the spectral index comes out as \cite{Li:2013hga}
\be
n_s = 4 -  2 \nu = 1 - 2 \Delta \frac{\ep}{(\ep -1)}, \label{eq:spectrum}
\ee
where we did not have to make any approximations (this was overlooked in \cite{Li:2013hga}). When the two exponents $b$ and $c$ in the original Lagrangian \eqref{eq:Lagrangian} are equal, we obtain an exactly scale-invariant spectrum, $n_s = 1.$ However, when $b$ and $c$ differ sightly, we obtain deviations from scale-invariance. Since we have $\ep > 3,$ the deviation from scale-invariance is always between $-3\Delta$ and $-2\Delta.$ Thus, if $b$ is larger than $c$ by about two percent, we obtain the central value $n_s = 0.96$ reported by the PLANCK team \cite{Ade:2013zuv}.

Using the large-scale expression for the mode functions \eqref{eq:modefunctionlate}, with $\nu$ given in \eqref{eq:spectrum}, we can find the time dependence of the original scalar field fluctuation $\de \chi:$
\be \label{eq:chilinear}
\de \chi = e^{\frac{b}{2}\phi} \frac{v_s}{a} \propto (-a \t)^{\frac{b}{c}}(-\t)^{\h-\nu}\frac{1}{a} = constant.
\ee
Thus $\de \chi$ tends to a constant on large scales, irrespective of the values of $b$ and $c,$ implying that this solution is stable for any values of the spectrum. The fact that $\de \chi$ is precisely constant (in the large-scale limit) will also have implications for the non-Gaussian corrections.


\indent {\it (No) Non-Gaussianity During The Ekpyrotic Phase.} In recent years, non-Gaussian corrections have become accessible to experiments and are playing an increasingly important role in discriminating between different cosmological scenarios. In the present paper, we will limit ourselves to calculating the bispectrum of the perturbations, leaving a calculation of higher $n$-point functions to future work. 

Since our model does not contain kinetic terms with more than two derivatives, it is immediately clear that non-Gaussianities of ``equilateral'' and ``orthogonal'' type are not going to get generated \cite{Langlois:2008qf}. Thus we can focus on evaluating the $3$-point function of ``local'' shape, which amounts to calculating the second order term in a Taylor series expansion (in real space) of the perturbations. We first evaluate the local bispectrum of the entropy perturbations. To this end, we expand the equation of motion for $\chi,$ Eq. \eqref{eq:chieom}, to second order. In doing so, one has to keep in mind the definitions of the perturbed scalar fields at second order. In comoving gauge, which is the gauge we will work in from now on, they are given by \cite{RenauxPetel:2008gi}
\be
\de\phi^{(2)} = - \frac{\de s^{(1)}}{2} \left(  \frac{\de s^{(1)\prime}}{\phi'}+\frac{b}{2}\de s^{(1)} \right), \, \de \chi^{(2)} = e^{\frac{b}{2}\phi} \de s^{(2)},
\ee
where we have indicated the perturbative order by a superscript. Since the potential does not depend on $\chi,$ and since the background satisfies $\dot\chi=0,$ it is then clear that the perturbed equation of motion will be identical to that at first order, Eq. \eqref{eq:deltaseom_2}. There arises no source term for the second-order entropy perturbation $\de s^{(2)},$ and we have the solution
\be \de s^{(2)} = 0. \ee
Hence no intrinsic non-Gaussianity is generated for the entropy perturbations! This is in contrast with the standard entropic mechanism, where the entropy perturbations develop significant local non-Gaussian corrections \cite{Creminelli:2007aq,Buchbinder:2007at,Lehners:2007wc,Lehners:2008my,Lehners:2013cka}.

The non-Gaussianities that are constrained by CMB measurements are those of the curvature perturbations $\zeta$, defined as local perturbations in the scale factor,
\be
\fd s^2 = - \fd t^2 + a(t)^2 e^{2 \z(t,x^i)} \fd x^i \fd x_i. \label{eq:zeta}
\ee 
During the ekpyrotic phase, curvature perturbations at linear order are not amplified \cite{Tseng:2012qd,Battarra:2013cha}, nor are they sourced by the entropy perturbations. However, at second order curvature perturbations can be sourced by the entropy fluctuations \cite{Lehners:2007wc}. To evaluate this contribution, we have to study the evolution equation for the curvature perturbations. A useful expression is provided by the following equation (which can be obtained by a trivial extension of known derivations \cite{Lyth:2004gb,Buchbinder:2007at,Lehners:2009qu} to the case of having a non-flat metric in field space), valid on large scales and expressed in comoving gauge ($\de \rho = 0$), 
\be \label{eq:curvls}
\dot{\z}= \frac{2 {H}  \de V}{\dot{{\phi}}^2 - 2 \de V }.
\ee
This remarkably simple equation is valid to all orders in perturbation theory. During the ekpyrotic phase, at second order and in conformal time, Eq. \eqref{eq:curvls} becomes 
\be 
\z^{(2)\prime} = -\frac{{\cH}a^2V_{,\phi} }{\phi^{\prime 2}} \de s^{(1)} \left(  \frac{\de s^{(1)\prime}}{\phi'}+\frac{b}{2}\de s^{(1)} \right).
\ee
Using the late-time/large-scale expression for the entropy mode functions \eqref{eq:modefunctionlate}, we obtain
\be
\frac{\de s^{(1)\prime}}{\phi'}+\frac{b}{2}\de s^{(1)} = \frac{1}{a}\left[ \frac{v_s'}{\phi'} - \frac{\cH v_s}{\phi'} +\frac{b}{2} v_s \right] =0.
\ee
Thus, amazingly enough, at second-order there is also no curvature perturbation generated during the ekpyrotic phase. This can be understood heuristically from the fact that the linearized solution is given by  $\de \chi^{(1)} = constant:$ if one thinks of the perturbation in the potential $\de V^{(2)}$ at second order (in Eq. \eqref{eq:curvls}) as a linear perturbation around the linearized solution, then it is clear that this vanishes in the same way as the linear perturbation vanishes around the constant $\chi$ background solution.

\indent {\it Time-varying equation of state.}
In order to have a successful model of the early universe, at some point the ekpyrotic phase must come to an end, which is most easily achieved if $\ep$ diminishes during the ekpyrotic phase. Thus, we are led to extend our model by allowing $\ep$ to be a slowly-varying function of time. Expressing the change in $\ep$ in terms of scale-factor ``time'' $\fd N = \fd \ln a,$ with $\frac{\fd}{\fd t}= H \frac{\fd}{\fd N},$ we can derive the following relations, valid in the large $\ep$ limit (see also \cite{Lehners:2007ac})
\bea \label{eq:a2V,phi-eplarge}
\phi' &\approx&  \tau^{-1} \sqrt{ \frac{2}{\ep}} \left( 1 + \frac{1}{\ep} + \frac{ \ep_{,N}}{\ep^2} + {\cal O}\left(\frac{1}{\ep^2}\right) \right), \\ a^2 V_{,\phi} &\approx& \tau^{-2} \frac{1}{\sqrt{2\ep}} \left( 2 - \frac{2}{\ep} + 3 \frac{\ep_{,N}}{\ep^2} + {\cal O}\left(\frac{1}{\ep^2}\right) \right).
\eea
The spectral index can then be approximated by
\be
n_s -1 = -2 \Delta \left(1+\frac{1}{\ep}\right) - \frac{7}{3} \frac{\ep_{,N}}{\ep^2} + {\cal O}\left(\frac{1}{\ep^2}\right). \label{eq:spectrumnonconstant}
\ee
As $N$ decreases during the ekpyrotic phase, a decreasing $\ep$ implies $\ep_{,N}>0,$ which shifts the spectrum of the perturbations slightly to the red. For the most symmetric case where originally $b=c,$ such a correction can naturally induce the observed small red tilt \cite{Ade:2013zuv}. 

The calculation of non-Gaussianities also changes somewhat, as we now obtain
\be
\frac{\de s^{(1)\prime}}{\phi'}+\frac{b}{2}\de s^{(1)} = -\frac{1}{6}\frac{v_s}{a}\sqrt{\frac{\ep}{2}} \left( \frac{\ep_{,N}}{\ep^2} +{\cal O}\left(\frac{1}{\ep^2}\right)\right).
\ee
Then there exists a small source term for the second-order curvature perturbation during the ekpyrotic phase,
\be
\z^{(2)\prime} =-\frac{1}{12} \frac{v_s^2}{a^2}\frac{1}{(-\t)}\left(\frac{\ep_{,N}}{\ep^2}+{\cal O}\left(\frac{1}{\ep^2}\right)\right).
\ee
If we approximate the time dependence of the fluctuation modes $v_s/a \propto 1/\t$ then we can easily perform the integral, obtaining
\be
\z_{ek-end}^{(2)}= -\frac{1}{24} (\de s_{ek-end})^2 \left(\frac{\ep_{,N}}{\ep^2}+{\cal O}\left(\frac{1}{\ep^2}\right)\right), \label{eq:zetaek}
\ee
where the subscript $_{ek-end}$ refers to the end of the ekpyrotic phase. We can see that the coefficient of $(\de s_{ek-end})^2$ is tiny, of ${\cal O}(10^{-2})$ at most for realistic cases. Thus, even with a time-varying equation of state, the non-minimal entropic mechanism generates perfectly Gaussian entropy perturbations and essentially vanishing curvature perturbations over the course of the ekpyrotic phase.


\indent {\it The Final Curvature Perturbations.} What we observe in the cosmic background radiation are not entropy perturbations, but rather the observed temperature fluctuations stem directly from curvature perturbations. Thus, for our model to be viable, we must ensure that the entropy perturbations can get converted into curvature fluctuations. Several possibilities for such a conversion process have already been discussed in the literature. 

The first is most easily explained by rewriting Eq. \eqref{eq:curvls} in terms of $\dot\theta = V_{,s}/\dot\phi,$ which represents the rate of change of the angle of the trajectory in scalar field space \cite{Gordon:2000hv}. At linear order one obtains
\be
\dot\zeta = \frac{2H}{\dot\phi} \dot\theta \de s = \sqrt{\frac{2}{\ep_c}} \dot\theta \de s, \label{eq:zetadotep}
\ee
where $\ep_c$ denotes the value of $\ep$ during the conversion process. This equation illustrates that whenever the background trajectory bends, curvature perturbations are generated. Since there is no $k$-dependence in Eq. \eqref{eq:zetadotep}, their spectrum will be identical to that of the entropy perturbations that source them, and thus will be given by Eq. \eqref{eq:spectrum} or \eqref{eq:spectrumnonconstant}. A bending could either occur at the end of the ekpyrotic phase \cite{Koyama:2007ag,Buchbinder:2007ad}, or during the subsequent kinetic phase where the ekpyrotic potential has become unimportant \cite{Lehners:2007ac}. For these two possibilities, we can estimate the amplitude of the curvature perturbation by approximating $\ep_c$ and $\de s$ as constants over the time of the conversion, and assuming a total bending angle of about $1$ radian, $\Delta \theta \approx 1,$ giving
$ \zeta_{final} \approx \frac{1}{\sqrt{\ep_c}}\de s_{ek-end}$ and leading to a power spectrum
\be
P_\zeta = \frac{k^3}{(2\pi)^2} \langle \zeta_{final}^2 \rangle \approx \frac{(\ep -1)^2}{\ep_c (\ep -3)} \frac{V_{ek-end}}{(2\pi)^2}, 
\ee  
where $V_{ek-end}$ corresponds to the energy scale of the deepest point in the potential. Unless the fast-roll parameter $\ep$ during the ekpyrotic phase is very close to $3$, this implies that (just as for the standard entropic mechanism) the potential has to reach the grand unified scale $V_{ek-end} \approx (10^{-2} M_{Pl})^4$ in order for the curvature perturbations to have an amplitude in agreement with the observed value of about $2\times10^{-9}$ \cite{Ade:2013zuv}. 

Another possibility, of a somewhat different character, is that conversion can occur at the bounce itself via the process of modulated (p)reheating \cite{Battefeld:2007st}. The idea here is that at the bounce massive matter particles can be copiously produced, with their subsequent decay into ordinary fermions being modulated by a coupling function $h(\de s).$ As shown in \cite{Battefeld:2007st}, the amplitude of the resulting perturbations is proportional to $h_{,s}/h,$ and thus all predictions depend on the ability to derive the precise form of $h(\de s)$ in a realistic setting. This conversion model has the property that no bending of the trajectory need to occur before the bounce.



The final non-Gaussianity in the curvature perturbation will depend on the non-linearity of the conversion process, and thus the final answer will be model-dependent. Nevertheless, we can get a general idea of the magnitude of the resulting non-Gaussianities by re-writing Eq. \eqref{eq:curvls} up to second order in yet another form,
\be
\zeta = \int \fd N \left[ \frac{2\de V}{\dot\phi^2} + \left( \frac{2\de V}{\dot\phi^2}  \right)^2 \right].
\ee
We may expect the conversion process to occur over about one e-fold of evolution of the universe. If the conversion process is {\it efficient}, i.e. if $\de V/\dot\phi^2$ does not vary much and in particular does not change sign over this e-fold of conversion, then we will have $\zeta^{(2)} \approx (\zeta^{(1)})^2$ and
\be
f_{NL}^{local} \equiv \frac{5}{3} \frac{\z^{(2)}}{(\z^{(1)})^2} \sim {\cal O}(1).
\ee
In this case the primordial local non-Gaussianity is of the same order of magnitude as the non-linear evolution that takes place between the surface of last scattering and now \cite{Pitrou:2010sn,Komatsu:2010hc}. Moreover, the contribution from the ekpyrotic phase in the case of non-constant $\ep,$ given in Eq. \eqref{eq:zetaek}, is then completely negligible. We can compare these expectations with the observations from the PLANCK satellite, which measured $f_{NL}^{local} = 2.7 \pm 5.8$ at the $1\sigma$ level \cite{Ade:2013ydc}. Thus, in this case our model is in perfect agreement with observations. 

Interesting questions for future research are to investigate under what conditions the conversion process is indeed efficient, and how the conversion process may best fit together with the bounce dynamics. What the non-minimal entropic mechanism ensures is that the starting point is ideal in that the perturbations coming out of the ekpyrotic phase have an essentially vanishing bispectrum.


\indent {\it Discussion.} As a model of the early universe, an ekpyrotic phase employing the non-minimal entropic mechanism has many compelling features. Contrasting with the standard entropic mechanism, we note that: 1. The initial conditions problem is markedly improved here, since the potential is everywhere stable. 2. The equation of state can be small $\ep \gtrsim 3,$ since near scale-invariance of the spectrum results from $b \approx c,$ regardless of the value of $c.$ 3. The entropy perturbations generated during the ekpyrotic phase have an exactly vanishing bispectrum. 

Contrasting with single-field plateau models of inflation, we note that: 4. The measure problem is essentially absent, since the ekpyrotic smoothing phase proceeds almost entirely at a very small Hubble rate \cite{Johnson:2011aa,Lehners:2012wz}. 5. If the process of converting entropic into curvature perturbations is efficient, the resulting curvature perturbations will have a small bispectrum, with non-Gaussianity parameters practically indistinguishable from those in single-field inflation, $f_{NL}^{equil} = f_{NL}^{ortho}=0$ and $f_{NL}^{local}={\cal O}(1).$ 6. In contrast to inflation, but just as in other ekpyrotic models, no large-amplitude primordial gravitational waves are produced \cite{Boyle:2003km,Baumann:2007zm}. Detecting these would in fact be the best way to falsify the model.

Thus we find that rather than adding an infinity of universes that are all physically different from each other, like eternal inflation does, the present model explains the current cosmological data with the addition of one scalar field and more involved dynamics. This appears to us to be an acceptable price to pay.


\acknowledgements 
We would like to thank Paul Steinhardt for his careful reading of the manuscript, and Anna Ijjas, Justin Khoury and Burt Ovrut for discussions. JLL thanks the Princeton Center for Theoretical Science for hospitality while this work was being completed. AF and JLL gratefully acknowledge the support of the European Research Council in the form of the Starting Grant No. 256994 ``StringCosmOS''.


\bibliographystyle{apsrev}
\bibliography{NonMinimalEntropicBib}

\end{document}